\documentstyle[11pt,newpasp,twoside]{article}

\input psfig.tex

\def\etal{{\em et~al.\ }}
\def\aa #1 #2 {A\&A, #1, #2}
\def\aas #1 #2 {A\&AS, #1, #2}
\def\acm #1 #2 {ACM-Trans Math Software, #1, #2}
\def\ada #1 #2 {Ann Astrophys, #1, #2}
\def\agabstr #1 #2 {Astr Ges Abstr Ser, #1, #2}
\def\aj #1 #2 {AJ, #1, #2}
\def\anach #1 #2 {Astr Nachr, #1, #2}
\def\apj #1 #2 {ApJ, #1, #2}
\def\apjl #1 #2 {ApJL, #1, #2}
\def\apjs #1 #2 {ApJS, #1, #2}
\def\araa #1 #2 {ARAA, #1, #2}
\def\apss #1 #2 {ApSpaceS, #1, #2}
\def\celmech #1 #2 {Cel Mech, #1, #2}
\def\esom #1 #2 {ESO Messenger, #1, #2}
\def\fundcp #1 #2 {FunCosP, #1, #2}
\def\jcp #1 #2 {J Comp Phys, #1, #2}
\def\jfm #1 #2 {J Fluid Mech, #1, #2}
\def\jmp #1 #2 {J Math Phys, #1, #2}
\def\ma #1 #2 {Mitt Astr Ges, #1, #2}
\def\mn #1 #2 {MNRAS, #1, #2}
\def\nat #1 #2 {Nat, #1, #2}
\def\obs #1 #2 {Observatory, #1, #2}
\def\pasj #1 #2 {PASJ, #1, #2}
\def\pasp #1 #2 {PASP, #1, #2}
\def\phyr #1 #2 {PhysRep, #1, #2}
\def\physd #1 #2 {Physica D, #1, #2}
\def\rpp #1 #2 {RepProgPhys, #1, #2}
\def\ssr #1 #2 {Sp Sci Rev, #1, #2}
\def\spose#1{\hbox to 0pt{#1\hss}}
\def\lta{\mathrel{\spose{\lower 3pt\hbox{$\mathchar"218$}}
     \raise 2.0pt\hbox{$\mathchar"13C$}}}
\def\gta{\mathrel{\spose{\lower 3pt\hbox{$\mathchar"218$}}
     \raise 2.0pt\hbox{$\mathchar"13E$}}}

\def\=#1{\overline{#1}}

\def\lvplot{($l,v$) diagram}

\def\rcr{R_{\rm CR}}

\def\phibar{\varphi_{\rm bar}}

\def\RSUN{R_0}

\def\rsun{{\rm\,R_0}}

\def\deg{^\circ}             

\def\kms{{\rm\,km\,s^{-1}}}

\def\pc{{\rm\,pc}}
\def\kpc{{\rm\,kpc}}

\def\msun{{\rm\,M_\odot}}

\def\rcr{R_{\rm CR}}

\def\phibar{\varphi_{\rm bar}}

\def\lvplot{($l,v$) diagram}

\def\RSUN{R_0}

\begin{document}

\title{Structure and Mass Distribution of the Milky Way Bulge and Disk}
\author{Ortwin Gerhard}
\affil{Astronomisches Institut, Universit\"at Basel,\\Venusstr.~7, 
CH-4102 Binningen, Switzerland}

\begin{abstract}
This article summarizes the structural parameters of the Galactic
bulge and disk, and discusses the interpretation of the bulge
microlensing observations and the determination of the Milky Way's
luminous mass from the terminal velocity curve and the Oort limit.
The bulge is a rotating bar with corotation radius around 4 kpc.  The
NIR disk has a short scale-length.  The measured surface density of
the local disk is in good agreement with the prediction of a maximum
NIR disk model for the Milky Way.  The preliminary new value for the
microlensing optical depth of clump giant sources is within
$1.7\sigma$ of the prediction by the maximum NIR disk model, while the
optical depth for all sources is still significantly higher. These
results imply that cold dark matter cannot dominate inside the
solar radius.
\end{abstract}

\section{Introduction}
\label{secintro}

The Milky Way is a normal galaxy seen under a magnifying glass: the
ages, metallicities, and kinematics of its stars can be observed in
much greater detail than in any other galaxy. Combined with a model
for the large-scale structure and potential, dynamical models for
specific stellar populations can be constructed which link our local
neighbourhood with distant parts of the Galaxy. Thus we can study in a
unique way the imprints of the evolution and formation processes that
shaped the Milky Way, and presumably other galaxies as well.

As is well-known, our viewpoint from within the Galactic disk also has
some drawbacks, in that many of the Milky Way's large scale properties
have been difficult to obtain. Only in the last decade has it become
possible to map the old stellar population in the Galactic bulge and
inner disk, using infrared observations to penetrate the dust layers
in the disk. Models are needed to interpret these data and link them
to gas kinematical and other observations. Two important results are
that the inner Galaxy contains a rotating bar/bulge, and that the
scale-length of the disk is significantly shorter than previously
assumed.

Microlensing surveys have now found hundreds of events towards the
Galactic bulge. From such data and from local stellar-kinematic
measurements, direct constraints on the mass of the Milky Way's bulge
and disk can be obtained, which are difficult to come by in external
galaxies, and which can be used to test cosmological models of dark
matter halo and galaxy formation.

\section{Structure of the Galactic bulge and disk}

The best evidence for a rotating bar in the inner Galaxy comes from
maps of the NIR light distribution, from star counts along various
lines-of-sight, from the non-circular motions observed in the atomic
and molecular gas, and from the large optical depth to microlensing
towards the bulge. For a more extensive review than given here see
Gerhard (1999, G99). All length scales given below are for a
Sun-center distance $\RSUN=8\kpc$.

Several models for the distribution of old stars in the inner Galaxy have
been based on the COBE/DIRBE NIR data. These data have complete sky coverage
but relatively low spatial resolution, they must still be `cleaned' for
residual dust absorption, and they contain no distance information, so
deprojection is not straightforward. The cleaned data show that the 
bulge is brighter and more extended in latitude $b$ at positive
longitudes $l$ than at corresponding $-l$, except for a region
close to the center where the effect is reversed (Weiland \etal 1994,
Bissantz \etal 1997). The asymmetry is strongest around $|l|\simeq10\deg$.
These signatures are as expected for a barred bulge with its long
axis in the first quadrant (Blitz \& Spergel 1991). The region of 
reversed asymmetry at small $|l|$ argues for a bar rather than a lopsided
light distribution; see also Sevenster (1999).

Dwek \etal (1995) and Freudenreich (1998) fitted parametric models to
the cleaned DIRBE data, assuming specific functional forms for the
barred bulge and excluding low-latitude regions from the fit. In this way the
derived outer bulge properties are less likely to be influenced by the
disk light, but the inner bulge and disk profiles are ill-determined.
Non-parametric models were constructed by Binney, Gerhard \& Spergel
(1997, BGS), using a Lucy algorithm based on the assumption of strict
triaxial symmetry, and by Bissantz \& Gerhard (2000), using a
penalized maximum likelihood algorithm. The new model by Bissantz \&
Gerhard includes a spiral arm model and maximizes eightfold symmetry
only for the remaining luminosity distribution.  In this model
the bulge-to-disk ratio in NIR luminosity is about $20\%$, similar
to the value given by Kent, Dame \& Fazio (1991). Other bar and disk
properties from the COBE models are summarized below.  Physical models
for the COBE bar can be found for a range of bar orientation angles,
$15\deg \lta \phibar \lta 35\deg$, where $\phibar$ measures the angle
in the Galactic plane between the bar's major axis at $l>0$ and the
Sun-center line. $\phibar$ must therefore be determined from other
data; see also Zhao (2000).

The bar is also seen in starcount observations in inner Galaxy fields.
Stanek \etal (1997) analyzed reddening-corrected apparent magnitude
distributions of clump giant stars in 12 OGLE fields. The small
intrinsic luminosity spread ($\sim$ 0.2-0.3 mag) makes these stars
good distance indicators.  The peak of the distribution is brighter at
$l>0$ where the line-of-sight passes through the near side of the
bar. These fields cover only a small fraction of the sky, but fitting
a parametric model constrains the bar orientation angle as well as the
axis ratios and density profile. Nikolaev \& Weinberg (1997)
reanalyzed the IRAS variable population in a similar spirit; here the
distance information comes from the known range of AGB star
luminosities.  NIR starcounts have also shown longitudinal asymmetries
due to the bar (Unavane \& Gilmore 1998). L\'opez-Corredoira \etal
(1997, 2000) and Hammersley \etal (1999) have modelled the Two Micron
Survey Starcounts (mostly bright K and M giants) in several strips
across the bulge. Structural information on the bulge and disk can be
derived from these data together with a model for the bright-star
luminosity function.  Ongoing work on deeper surveys (ISOGAL, DENIS,
2MASS) will provide important new information on the old stellar
population in the inner Galaxy; a preview of results is given in van
Loon (2000).

Modelling the HI and CO \lvplot s provides information on the
gravitational potential of the bar and disk. Several recent gas flow
models have followed complementary approaches: Englmaier \& Gerhard
(1999) modelled the gas flow in the potential of the rotating COBE bar
of Binney, Gerhard \& Spergel (1997), Fux (1999) determined a
best-fitting N-body--SPH model from a sequence of ab initio
simulations, and Weiner \& Sellwood (1999) considered gas flows in a
sequence of analytic model potentials.  These simulations produce
\lvplot s with which many features seen in the observed \lvplot s may
be qualitatively understood, such as the 3 kpc arm, the non-circular
velocities around the end of the bar, the cusped-orbit shock
transition and inner $x_2$-disk, the molecular ring and the spiral arm
tangent locations. However, no model as yet provides a satisfactory
quantitative account of the Galactic gas disk. A fuller discussion can
be found in G99.

The following subsections contain my best summary of the main
bar and disk parameters from this and other work.

{\noindent\bf Corotation radius:} This is the most important bar
parameter. The gas-dynamical simulations of Englmaier \& Gerhard
(1999) and Fux (1999) agree in their interpretation of the 3 kpc arm
as one of the lateral arms close to the bar, placing it inside
corotation.  Sevenster (1999) argues that the 3 kpc arm is part of an
inner ring, which would also place it slightly inside the corotation
radius $\rcr$.  The main Galactic spiral arms outside $\rcr$, on the
other hand, imply an upper limit for $\rcr$, but this is more
model-dependent. The gas-dynamical models thus give a range of $3\kpc
\lta \rcr \lta 4.5\kpc$; a corresponding resonance diagram is shown in
G99. Dehnen (2000) has interpreted features in the local stellar
velocity distribution as due to the outer Lindblad resonance with the
bar, resulting in $\rcr =0.55\pm0.05 R_0$, near the upper end of
the range from the gas-dynamical models. While the match to the
Hipparchos data appears convincing, it is not clear whether and why
this way of determining $\rcr$ should work, because the bar is too
weak to excite the spiral arms between $\rcr$ and the solar radius,
and these spiral arms should dominate the local quadrupole moment
(Englmaier \& Gerhard 1999).

{\noindent\bf Bar orientation} From the integrated light alone,
physically reasonable models can be found for $15\deg \lta \phibar
\lta 35\deg$. Starcount models give values between
$\phibar=12\pm6\deg$ (L\'opez-Corredoira \etal 2000) and
$20-30\deg$ (Nikoalev \& Weinberg 1997, Stanek \etal 1997).  The
models of Bissantz \& Gerhard (2000) for the DIRBE L-band data, when
additionally constrained by the clump giant apparent magnitude
distributions of Stanek \etal (1997), give an optimal $\phibar=15\deg$
(Figure 1), but $\phibar\simeq15-25\deg$ is within the uncertainties.
The gas-dynamical models and the orbit analysis of Binney \etal (1991)
are also compatible with $15\deg \lta \phibar \lta 35\deg$, depending
on whether the emphasis is on the peak in the terminal velocity curve,
the arm morphology, or the magnitude of the non-circular motions near
the 3 kpc arm. Finally, microlensing observations favour
$\phibar\sim15\deg$ (Zhao \& Mao 1996, \S3).  Thus a good working
value is $\phibar=20\deg$. Not consistent with this appear to be the
bar model of Hammersley \etal (2000), which is based on the
identification of a region of strong star formation at $l=27\deg$ with
the nearer end of the bar, and the star count results reported in van
Loon (2000), which place the near end of a 1.4 kpc size bar at
negative longitudes.

{\noindent\bf Bar length:} Models based on the DIRBE NIR maps find
the end of the bar around $R_{GC}=3.2\pm0.3\kpc$, when $\phibar\simeq
20\deg$ (Freudenreich 1998, BGS, Bissantz \& Gerhard 2000). This is
consistent with with the OH/IR stars (Sevenster 1999), IRAS variables
(Nikolaev \& Weinberg 1997), and the range of $\rcr$ above for a fast
bar, while other starcount models use exponential or Gaussian density
distributions with shorter scale-lengths.

{\noindent\bf Bar axis ratios:} The parametric DIRBE models give axial
ratios of about 10:3-4:3. This is in agreement with the new
non-parametric model of Bissantz \& Gerhard (2000), whereas BGS had
found 10:6:4 without taking into account the spiral arms. The
starcount models give 10:4:3 (Stanek \etal 1997) and 10:5.4:3.3
(L\'opez-Corredoira \etal 2000).  Thus there is good overall agreement
at around 10:4:3.

\begin{figure}
\centerline{
  \psfig{figure=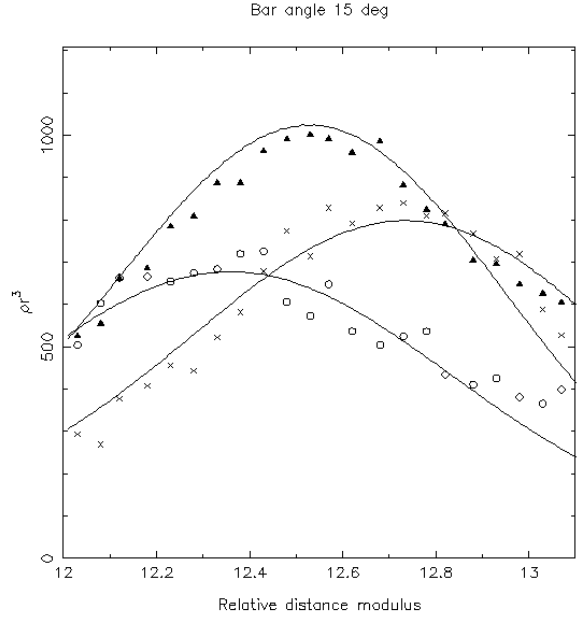,width=8.5cm}}
\caption{\small Apparent magnitude distribution of clump giant stars
in three fields observed by Stanek \etal (1997) with superposed scaled
model predictions from Bissantz \& Gerhard (2000) for their model with
$\phibar=15\deg$. The model curves have been normalized and shifted
along the abscissa such that the $l=5.5\deg$ and $l=-4.9\deg$ peaks
match best the locations of the observed peaks (circles and crosses,
respectively) . Note that the relative number, width, and
asymmetry of the observed distributions in these fields and in Baade's
window (triangles) are matched well.  }
\end{figure}

{\noindent\bf Disk scale-length:} In the integrated NIR the radial
exponential disk scale $R_D$ is significantly shorter than in the
optical; numerical values are around $2.5\kpc$ (Freudenreich 1998,
BGS) or somewhat shorter ($2.1\kpc$, Bissantz \& Gerhard 2000).
Hammersley \etal (1999) report satisfactory agreement of their NIR
counts with a model with $R_d=3.5\kpc$, but have not tested other
values of $R_D$.  L\'opez-Corredoira \etal (2000) find that
$R_d=3.0\kpc$ is too short to describe the NIR TMSS counts
well. Earlier starcount models (Robin \etal 1992, Ortiz \& L\'epine
1993) favour a short disk scale, $R_d=2.5\kpc$. It is not clear what
causes the differences between the various starcount models, and
between the TMSS starcounts and the integrated NIR light.  There may
be some interplay between the disk scale-length, the bulge profile,
and the spiral arm luminosity distribution. Further work with
spatially complete data will be needed to assess this.

{\noindent\bf Spiral structure:} Spiral arms are difficult to
delineate from our viewpoint within the disk. Modelling their
($l,v$)-traces directly is complicated by non-circular motions (Burton
1973). The most reliable information is available for the arm tangent
directions of the spiral arms in the gas and young stars (see review
in Englmaier \& Gerhard 1999). The arm segments connecting the tangent
points are more difficult to determine because of uncertain tracer
distances (but see Dame \etal 1986). The most widely used spiral model
is a four-armed pattern following Georgelin \& Georgelin (1976).  Such
a pattern also appears most consistent with gas-dynamical models for
the Milky Way (Englmaier \& Gerhard 1999, Fux 1999). Much less is
known about spiral arms in the old Galactic disk stars (we know from
external galaxies that these need not follow the blue arms). NIR plane
profiles do not clearly show all the spiral arm tangents seen in the
young component (Drimmel \& Spergel 2000), and the contribution of
young supergiant stars to those that are seen has not been
evaluated. Drimmel \& Spergel (2000) conclude that both a two-armed
and a sheared (relative to the gas and young stars) four-armed pattern
are consistent with the DIRBE data. The best models of Fux (1999)
contain four stellar spiral arms.

\section{Microlensing towards the Galactic bulge}

Several hundred microlensing events have now been observed towards the
Galactic bulge. These observations give information about the
integrated mass density towards the survey fields as well as about the
lens mass distribution. The most robust observable is the total
optical depth averaged over the observed fields, $\tau$.  Early
measurements gave surprisingly high values $\tau_{-6}\simeq 2-4$
(Udalski \etal 1994, Alcock \etal 1997), where
$\tau_{-6}\equiv\tau/10^{-6}$. For clump giant sources only, which do
not suffer from blending problems, Alcock et al.\ (1997) inferred
$\tau_{-6} = 3.9^{+1.8}_{-1.2}\times 10^{-6}$ from 13 events centered
on $(l,b) = (2.55\deg, -3.64\deg)$.  Using a difference image analysis
(DIA), Alcock et al.\ (2000a) recently measured
$\tau_{-6}=2.43^{+0.39}_{-0.38}$ for all sources from 99 events
centered on $(l,b) = (2.68\deg, -3.35\deg)$, and from this measurement
deduced for the same direction $\tau=(3.23\pm0.5)\times10^{-6}$ for
bulge sources only. Finally, in a preliminary analysis of 52 clump
giant sources in 77 Macho fields, Popowski \etal (2000) found a lower
$\tau_{-6}=2.0\pm0.4$ centered on $(l,b)=(3.9\deg,-3.8\deg)$.

It has long been known that axisymmetric models predict
$\tau_{-6}\simeq 1-1.2$, insufficient to explain the quoted optical
depths (Kiraga \& Paczynski 1994, Evans 1994).  Models with a nearly
end--on bar as described in \S 2 enhance $\tau$ because of the longer
average line-of-sight from lens to source.  The maximum effect occurs
for $\phi\simeq \arctan(b/a)$ when $\tau_{\rm bar}/\tau_{\rm
axi}\simeq (\sin2\phi)^{-1}\simeq 2$ for $\phi=15\deg$ (Zhao \& Mao
1996).  In addition, $\tau$ increases with the mass and the length of
the bar/bulge.

Even so, models based on barred mass distributions derived from Milky
Way observations (\S 2) typically give $\tau_{-6}\simeq 1-2$ (e.g.,
Zhao, Spergel \& Rich 1995, Stanek \etal 1997, Bissantz \etal 1997),
significantly less than most of the measured optical depths. The new
bar model of Bissantz \& Gerhard (2000) gives $\tau_{-6}=1.2$ for all
sources at the position of the DIA measurement and $\tau_{-6}=1.3$ for
clump giant sources at the centroid position given by Popowski \etal
(2000). The mass normalization of the disk and bulge in this model is
calibrated by assuming constant L-band mass-to-light ratio and by
matching the predicted gas flow velocities in a hydrodynamic
simulation to the Galactic terminal velocity curve.  As Fig.~1 shows,
the apparent magnitude distributions for clump giant stars predicted
by this model agree closely with those measured by Stanek \etal
(1997). Thus the model gives a good approximation to the distribution
of microlensing sources.  The quoted optical depths are therefore hard
to change unless one assumes that the mass distribution of the lenses
differs substantially from that of the sources.

Notice that the preliminary new MACHO clump giant optical depth is within
$1.7\sigma$ of the prediction of this bar model. If this clump giant
optical depth is confirmed, this would be an important step in
reconciling galactic structure and microlensing observations and would
enable us to start using more detailed microlensing observables as
constraints on Galactic models. On the other hand, the recent DIA
value is still some $3.2\sigma$ away from the model prediction.  While
the NIR model prediction could be slightly increased if the
mass-to-light ratio were not spatially constant, this is only a $\sim
20\%$ effect since limited by the terminal velocity curve (Bissantz
\etal 1997).  Recently, Binney, Bissantz \& Gerhard (2000) have used
general arguments to show that an optical depth for bulge sources as
large as derived by the MACHO collaboration from the DIA value is very
difficult to reconcile with measurements of the rotation curve and
local mass density, even for a barred model and independent of whether
mass follows light. To illustrate this, the extra optical depth
required corresponds to an additional mass surface density towards the
bulge of some $2000 \msun/\pc^2$, comparable to that in the
model of Bissantz \& Gerhard (2000). Is it possible that the DIA
measurement is still significantly affected by blending?

Independent of the resolution of this problem, these results have a
further important implication. For, determining the mass normalization
of the Bissantz \& Gerhard (2000) model from the terminal velocities
implicitly assumes a maximal disk. Because the predicted microlensing
optical depths are if anything low, as much microlensing matter is
needed within the solar circle as possible. On the other hand, little
of the halo dark matter causes microlensing (Alcock \etal 2000b), so
we can not afford significantly more dark mass inside the solar circle
in the form of CDM particles, say, than corresponding to this maximum
disk model.  Thus the microlensing results argue strongly
for a maximum disk.

\section{The maximum disk and local surface density of the Milky Way}

Englmaier \& Gerhard (1999) computed a number of gas flow models in
the gravitational potential of the NIR COBE model determined by BGS,
assuming a constant joint NIR mass-to-light ratio $\Upsilon_L$ for the
bulge and disk.  The actual value of $\Upsilon_L$ can be specified a
posteriori.  For a maximum disk model it is determined by fitting the
model terminal curves to the observed HI and CO terminal velocities.
Some of the models included an additional dark component to prevent
the outer rotation curve from falling, leading to higher circular
velocities at $R_0$.  It turns out that the fitted value $\Upsilon_L$
is insensitive to the circular velocity $V_0$ of the Sun implied by
the model: within this class of models, the mass of the disk and bulge
are fixed to within $\sim 10\%$. The luminous component in these
models accounts for the terminal velocities out to $|l|\simeq 45\deg$,
or $R\simeq 5.5\kpc\simeq 2R_D$ if $V_0=220\kms$, and out to $R_0$ if
$V_0=180\kms$.

These maximum disk models predict a surface mass density $\Sigma_\odot
= 45 \msun/\pc^2$ near the Sun at $\rsun=8\kpc$ for a local circular
velocity of $v_{0}=208-220\kms$. For comparison, the local surface
density of `identified matter' is $48\pm 9 \msun/\pc^2$ (Kuijken \&
Tremaine 1991, Flynn \& Fuchs 1994, Holmberg \& Flynn 2000).  Of this
about $23\msun/\pc^2$ is in gas and brown and white dwarfs, which
contribute most to the uncertainty.  That the observed and predicted
surface density approximately agree lends support to the conclusion
that the Galaxy indeed has a near--maximum disk; the combined
observational and model uncertainty is about $30\%$ in mass. A NIR
disk accounting for only $60\%$ of the rotation velocity at $2R_D$, as
advocated for spiral galaxies by Rix \& Courteau (1999), would have
only $\Sigma_\odot \simeq 16 \msun/\pc^2$. Turned around, if the NIR
disk and bulge are given the $\Upsilon_L$ value implied by the local
surface density measurement, then they account for the observed
terminal velocities in the inner Galaxy.  Compared to earlier
analyses, the main difference is the short disk scale--length
($2.5\kpc$ in the model of BGS; see also Sackett 1997) -- the Sun is
well beyond the maximum in the rotation curve from only NIR luminous
matter.

When a cored spherical halo is added to the maximum NIR disk, such as to
make the Galaxy's rotation curve approximately flat at $v_c=220\kms$,
the halo core radius comes out $R_c\simeq 15\kpc$. This shows that the
Milky Way's dark halo is not very strongly concentrated.  Integrating
the surface density of this halo between $z=\pm1.1\kpc$ gives
$\Sigma_{h,1.1}=16 \msun/\pc^2$.  Adding this to the surface density
of the old NIR disk and the thick disk ($\simeq 9\msun/\pc^2$, see
discussion in G99), the total is $\Sigma_{\rm NIR}+\Sigma_{\rm
th}+\Sigma_{h,1.1} = 71 \msun/\pc^2$, whereas the measured total
$\Sigma_{1.1} = 71\pm 6 \msun/\pc^2$ (Kuijken \& Gilmore 1991).  This
very good agreement is clearly better than one expects, given the
uncertainties in both numbers. Note that for smaller values of the
Galaxy's asymptotic rotation velocity, the required amount of halo
would be reduced; recall that for $v_c=180\kms$ the terminal velocity
curve can be fitted within the errors without any added halo (at the
prize of a falling rotation curve). A more detailed analysis and
comparison with cosmologically motivated Milky Way halos is clearly
needed.


\begin{references}
\small

\reference 
Alcock C., \etal, 1997, ApJ, 479, 119 

\reference
Alcock, C., et al.\ 2000a, ApJ, 541, 734

\reference
Alcock, C., et al.\ 2000b, ApJ, 542, 281

\reference
Binney, J.J., Bissantz N., Gerhard, O.E., 2000, ApJL, 537, L99

\reference
Binney, J.J., Gerhard, O.E., \& Spergel, D.N.\ 1997, MNRAS, 288, 365 (BGS)

\reference
Binney, J.J., Gerhard, O.E., Stark, A.A., et al., 1991, MNRAS, 252, 210

\reference
Bissantz, N., Englmaier, P., Binney, J.J., \& Gerhard, O.E.\ 1997, MNRAS,
{289}, {651}

\reference
Bissantz, N., \& Gerhard, O.E.\ 2000, MNRAS, to be submitted

\reference 
Blitz L., Spergel, D., 1991, ApJ, 379, 631                      

\reference 
Burton W.B., 1973, PASP, 85, 679

\reference 
Courteau S., Rix H.-W., 1999, ApJ, 513, 561

\reference 
Dame T.M., Elmegreen B.G., Cohen R.S., Thaddeus P., 1986, ApJ, 305, 892

\reference
Dehnen W., 2000, AJ, 119, 800

\reference
Drimmel R., Spergel D.N., 2000, ASP, in press, astro-ph/0008313

\reference 
Dwek E., \etal, 1995, ApJ, 445, 716

\reference
Englmaier, P. \& Gerhard, O.E.\ 1999, MNRAS, 304, 512

\reference 
Evans N.W., 1994, \apjl{437} {L31}                               

\reference 
Flynn C., Fuchs B., 1994, \mn{270} {471}                         

\reference 
Freudenreich H.T., 1998, ApJ, 492, 495                          


\reference
Fux R.\ 1999, \aa, 345, 787

\reference
Georgelin Y.M., Georgelin Y.P., 1976, A\&A, 49, 57

\reference
Gerhard O.E., 1999, in Galaxy Dynamics, ASP, 182, 307 (G99)

\reference 
Hammersley P.L., Cohen M., Garz\'on F., et al., 1999, \mn{308} {333}

\reference 
Hammersley P.L., Garz\'on F., Mahoney T.J., et al., 2000, \mn{317} {L45}

\reference
Holmberg, J., \& Flynn, C.\ 2000, MNRAS, 313, 209


\reference 
Kent S.M., Dame T.M., Fazio G., 1991, ApJ, 378, 131             

\reference 
Kiraga M., Paczy\'nski B., 1994, \apj 430 L101           

\reference 
Kuijken K., Gilmore G., 1991, \apj{367} {L9}             

\reference 
L\'opez-Corredoira M., Garz\'on F., Hammersley P.L., et al., 1997, 
        \mn{292} {L15}                      

\reference 
L\'opez-Corredoira M., Hammersley P.L., Garz\'on F., et al., 2000, 
        \mn{313} {392}

\reference 
Nikolaev S., Weinberg M.D., 1997, ApJ, 487, 885                 

\reference 
Ortiz R., L\'epine J.R.D., 1993, \aa{279} {90}           


\reference
Popowski P., et al., 2000, ASP, in press, astro-ph 0005466

\reference 
Robin A.C., Cr\'ez\'e M., Mohan V., 1992, \aa{265} {32}  

\reference 
Sackett P.D., 1997, ApJ, 483, 103                                   

\reference 
Sevenster M.N., 1999, MNRAS, 310, 629


\reference 
Stanek K.Z., et al.\ 1997, ApJ, 477, 163                        

\reference 
Udalski A. \etal, 1994, Acta Astron., 44, 165 

\reference 
Unavane M., Gilmore G., 1998, \mn{295} {145}             

\reference
van Loon J.T., 2000, ASP, in press, astro-ph/00009471

\reference 
Weiland J.L., \etal, 1994, \apjl{425} {L81}                     

\reference 
Weiner B., Sellwood J.A., 1999, \apj{524} {112}

\reference 
Zhao H.S., 2000, MNRAS, 316, 418

\reference 
Zhao H.S., Mao S., 1996, MNRAS, 283, 1197                        

\reference 
Zhao H.S., Spergel D.N., Rich R.M., 1995, ApJ, 440, L13  

\end{references}
\end{document}